\documentclass[showpacs,aps,amsmath,amssymb,twocolumn]{revtex4-2}

\usepackage[]{graphicx}
\usepackage{color}
\usepackage{bbold}
\usepackage{braket}
\usepackage[colorlinks=true, pdfstartview=FitV, linkcolor=blue, citecolor=red, urlcolor=black]{hyperref}
\newcommand{\be}{\begin{equation}}
\newcommand{\ee}{\end{equation}}
\newcommand{\ben}{\begin{eqnarray}}
\newcommand{\een}{\end{eqnarray}}
\newcommand{\bes}{\begin{subequations}}
\newcommand{\ees}{\end{subequations}}
\def\bal#1\eal{\begin{align}#1\end{align}}

\newcommand{\nn}{\nonumber\\}
\newcommand{\bfi}{\begin{figure}}
\newcommand{\efi}{\end{figure}}
\newcommand{\bc}{\begin{center}}
\newcommand{\ec}{\end{center}}

\newcommand{\arcsinh}{\mbox{arcsinh}}

\begin{document}
\title{Multi-kink braneworld configurations \\
in the scalar-tensor representation of $f(R,T)$ gravity}
\author{D. Bazeia$^{1}$, A. S. Lob\~ao Jr.$^{2}$ and Jo\~ao Lu\'is Rosa$^{3}$}

\affiliation{$^1$Departamento de F\'isica, Universidade Federal da Para\'iba, 58051-970 Jo\~ao Pessoa, PB, Brazil\\$^2$Escola T\'ecnica de Sa\'ude de Cajazeiras, Universidade Federal de Campina Grande, 58900-000 Cajazeiras, PB, Brazil\\$^3$Institute of Physics, University of Tartu, W. Ostwaldi 1, 50411 Tartu, Estonia}
\begin{abstract}
In this work we investigate the $f(R,T)$ brane in the scalar-tensor representation, where the solutions of the equations of motions for the source field engender topological defects with two-kink profiles. We use the first-order formalism to obtain analytical solutions for the source field of the brane and analyze how these solutions modify the structure of the auxiliary fields arising from the scalar-tensor representation of the theory. We found that when the model engenders two-kink solutions, the auxiliary fields are modified in order to allow for the appearance of an internal structure. In addition, the stability potential and zero mode also have their internal structure modified by two-kink solution.
\end{abstract}

\maketitle
\section{Introduction}

Modified gravity theories have been widely used to study the current cosmological scenario, considering that several opened questions in cosmology do not seem to obey the usual prescription formulated through the Einstein-Hilbert action. In this sense, various proposals for extensions of gravity have been presented in an attempt to clarify important issues such as inflation, dark energy, quintessence, local gravity constraints and others \cite{Clifton:2011jh}. 

It is known that gravity can be generalized in several ways. Popular extensions widely used in the literature are the $f(R)$ gravity \cite{Nojiri:2006ri,DeFelice:2010aj}, Gauss-Bonnet \cite{Nojiri:2005vv,Nojiri:2005jg}, teleparallel \cite{Cai:2015emx,BeltranJimenez:2019odq,Bahamonde:2021gfp}, cubic gravity \cite{Bueno:2016xff,Erices:2019mkd}, Galileon theory \cite{Nicolis:2008in,Leon:2012mt}, among others. Many of the ideas that emerged in generalized gravity have been used to study others intriguing theoretical questions as e.g. braneworld models, which are studied in the framework of five-dimensional gravity, where the extra dimension has infinite extension and the geometry is deformed by a real function; see, e.g., Refs. \cite{Randall:1999vf,Goldberger:1999uk,Skenderis,Csaki:1999mp,DeWolfe:1999cp,D,R1, E,Gremm:1999pj,Brito:2001hd, Bazeia:2003cv,Kobayashi:2001jd,Ee,Guerrero:2006gj, F, G, H,R2,R3} and references therein. In such contexts, modifying the formulation of the theory can interfere with the perturbation spectra of the solutions, altering the way the graviton is trapped in the brane \cite{Bazeia:2013uva,Bazeia:2015oqa,Bazeia:2015owa}.

Some of these generalized models raise new and interesting questions regarding braneworld scenarios. However, given the robustness and complexity of the modified field equations at a differential level, analytical solutions are commonly unattainable. Frequently, one restricts their analysis to simple particular forms of the theory in order to proceed with the investigation analytically, see e.g. \cite{Afonso:2007gc} where the $f(R)-$brane is investigated with a constant Ricci scalar, or \cite{Bazeia:2015owa} where analytical solutions are obtained for a simple $f(R,T)$ brane in the form $f(R,T)=R+T$, where $T$ is the trace of the stress-energy tensor.

A particularly interesting way of investigating generalized gravity is through dynamically equivalent representations which do not require the restriction to specific forms of the model. A representation that allows to deal with generalized braneworld scenarios is the scalar-tensor representation, where the extra scalar degrees of freedom of the theory are exchanged by auxiliary scalar fields. For example, the scalar-tensor representation of the $f(R,T)$ gravity introduced in Ref. \cite{Rosa:2021teg} was proved to be useful in a cosmological context \cite{Goncalves:2021vci,Goncalves:2022ggq} and was used in Refs. \cite{Rosa:2021tei,Rosa:2022fhl} to study $f(R,T)$ brane; in this context, two auxiliary scalar fields were introduced to carry the two scalar degrees of freedom associated with the arbitrary dependence of the function $f(R,T)$ in the Ricci scalar $R$ and trace of stress-energy tensor $T$. This formalism was also used to investigate a scenario where the brane engendered Cuscuton dynamics \cite{Rosa:2021myu}.

In the scalar-tensor representation for braneworld models, it is assumed that the solutions of the source fields are known. This guarantees the usual behavior of the warp factor and energy density of the model, while the solutions for the auxiliary fields are obtained via their own equations of motion. In this case, if we consider topological solutions with different behaviors as ansatz for the source fields, e.g. kink-compact or multi-kinks solutions, the structure of the auxiliary fields is changed, introducing new possibilities and leading to a better understanding of how generalizations of gravity modify the braneworld scenario.

In this perspective, we are particularly interested in understanding how the auxiliary fields of the scalar-tensor representation of the $f(R,T)$ brane are modified when dealing with two-kink solutions for the source field. To this purpose, we consider models that engender solutions with two-kink profiles. In particular, we investigate two interesting cases; the first one allowing us to have kink or two-kink solutions by adjusting a discrete parameter. This model is known as the $p$-model and was originally investigated in \cite{Bazeia:2003qt}. The second case is a theoretical model investigated in \cite{Bazeia:2005hu}. In this second model, the kink solution is distorted inducing the emergence of a two-kink profile. It is known that multi-kink solutions have been obtained in generalized braneworld models, see e.g. \cite{Xu:2022ori} where the authors obtain multi-kink brane solutions in the context of Gauss-Bonnet gravity. In this sense, the use of different solution profiles for the source field can introduce new possibilities in the study of generalized brane models.

In view of what has been presented, we organize this work as follows. Sec. \ref{formaism} provides the general formalism that describes the scalar-tensor representation of the $f(R,T)$ gravity and its application to the study of a five-dimensional brane. In this section we will also introduce the first-order formalism in order to obtain analytical solutions. In Sec. \ref{secStability} we investigate the linear stability of static solutions. In Sec. \ref{SpecificModels} we investigate models that engender two-kink solutions and how such solutions modify the auxiliary fields. In Sec. \ref{coments} we present the conclusions and perspectives for future work.

\section{Theory and framework}\label{formaism}

Let us start by describing the scalar-tensor representation of the $f(R,T)$ brane, where $R$ is the Ricci scalar and $T\equiv T_a{}^a$ is the trace of the stress-energy tensor $T_{ab}$. For this, let us assume a generalized action $S$ in five-dimensions in the form
\begin{equation}\label{actiongeo}
S=\int_\Omega\!\!\sqrt{|g|}\,d^5x\left[\frac{1}{4}f\left(R,T\right)-{\cal L}_s\right],
\end{equation}
where $\Omega$ is a five-dimensional spacetime manifold described by a set of coordinates $x^a$ and $g=det(g_{ab})$. Moreover, we using natural units and $4\pi G_5=1$, where $G_5$ is the gravitational constant. In this study, the Greek indices $\mu,\nu, ...$ range from $0$ to $3$ and Latin indices $a,b, ...$ range from $0$ to $4$. Furthermore, we will consider a standard Lagrangian density describing a scalar field $\chi$ as the source of the brane model, i.e.,
\begin{equation}\label{lagrangeStandar}
    {\cal L}_s=\frac12\nabla_a\chi\nabla^a\chi-V(\chi),
\end{equation}
where $\nabla_a$ are the covariant derivatives with respect to coordinates $x^a$ and $V(\chi)$ is the potential of the brane.

In this paper we are particularly interested in situations where $f_{RR}\,f_{TT}\neq f_{RT}^2$, where the indices $R$ and $T$ represent the derivatives of $f(R,T)$ with respect to these functions. This case was discussed in Ref. \cite{Rosa:2021teg}, where the authors showed that it is possible to construct a dynamically equivalent representation of the action \eqref{actiongeo} introducing two auxiliary scalar fields $\varphi$ and $\psi$ and an interaction potential $U\left(\varphi,\psi\right)$ as
\ben
\varphi=\frac{\partial f}{\partial R}\,,\qquad\qquad \psi=\frac{\partial f}{\partial T},
\een
\ben
U\left(\varphi,\psi\right)=-f\left(R,T\right)+\varphi R + \psi T.
\een

Thus, Eq. \eqref{actiongeo} can be rewritten as
\ben
\!\!\!\!S=\!\!\int_\Omega\!\!\!\sqrt{|g|}\,\left[\frac14\varphi R+\frac14\psi T-\frac14U\!\left(\varphi,\psi\right)
-{\cal L}_s\right]d^5x.\label{actionst}
\een
In this representation, the extra scalar degrees of freedom introduced by the $f(R,T)$ function are carried by the auxiliary fields $\varphi$ and $\psi$.

Let us now obtain the equations of motion. Varying Eq. \eqref{actionst} with respect to the metric $g_{ab}$ we obtain the modified field equations as
\begin{equation}
\begin{aligned}
\!\!\!\!&-\frac{1}{2} g_{ab}\left(\varphi R-U\right)+\varphi R_{ab}-\left(\nabla_a\nabla_b-g_{ab}\, \nabla_c\nabla^c\right)\varphi\\
\!\!\!\!& =2\, T_{ab}+\frac32\psi\nabla_a\chi\nabla_b\chi +\frac12\,g_{ab}\,\psi\, T ,\label{fieldst}
\end{aligned}
\end{equation}
where the stress-energy tensor $T_{ab}$ is given by
\begin{equation}\label{defTab}
    T_{ab}=\,\nabla_a\chi\nabla_b\chi -\frac12g_{ab}\nabla_c\chi\nabla^c\chi+g_{ab}V.
\end{equation}
Taking the trace of the above equation with the inverse metric $g^{ab}$ we obtain
\begin{equation}\label{defTraço}
    T=-\frac32\nabla_a\chi\nabla^a\chi+5V.
\end{equation}
On the other hand, taking a variation of Eq. \eqref{actionst} with respect to the source field $\chi$, the equation of motion for $\chi$ is
\ben\label{eq142}
\nabla_a\nabla^a\chi+V_\chi =-\,\frac{3}{4}\nabla_a\big(\psi\, \nabla^a\chi\big)-\frac{5}{4}\psi V_\chi,
\een
where $V_\chi=dV/d\chi$. Finally, taking the variation of Eq.~\eqref{actionst} with respect to $\varphi$ and $\psi$, we can also obtain the equations of motion for the auxiliary scalar fields as
\begin{equation}\label{eomphi}
    U_\varphi=R,\qquad\qquad U_\psi=T,
\end{equation}
where $U_\varphi=\partial U/\partial\varphi$ and $U_\psi=\partial U/\partial\psi$. 

Let us now assume the usual five-dimensional metric of brane models,
\begin{equation}\label{metricbrane}
    ds^2=e^{2A}\eta_{\mu\nu}dx^\mu dx^\nu-dy^2\,,
\end{equation}
where the four-dimensional Minkowiki metric $\eta_{\mu\nu}$ has a positive signature, i.e., $(+\,-\,-\,-)$. Furthermore, we will study static configurations where the scalar fields and the warp function depend solely on the extra dimension, i.e., $A=A(y)$, $\psi=\psi(y)$, $\varphi=\varphi(y)$ and $\chi=\chi(y)$. With this prescription we can write the non-vanish and independent components of the modified field equations given in Eq. \eqref{fieldst} as
\be
\begin{aligned}
&-6\varphi\big(A''+2A'^2\big)-6\varphi' A'-2\varphi''+U\\
&=2\chi^{\prime 2}+4V+\left(\frac32\chi^{\prime 2}+5V\right)\psi,\label{field1}
\end{aligned}
\ee
and
\be
\begin{aligned}
12\varphi A'^2+8\varphi' A'-U\!=\!2\chi^{\prime 2}\!-\!4V\!+\!\left(\frac32\chi^{\prime 2}\!-\!5V\!\right)\psi,\label{field2}
\end{aligned}
\ee
where a prime ($'$) denotes derivatives with respect to the extra dimension $y$. The equation of motion for $\chi$ in Eq. \eqref{eq142} becomes
\ben
\!\!\!\!\chi''\!+\!4A'\chi'\!=\!V_\chi\!-\!\frac{3}{4}\chi'\psi'\!-\!\left(\frac{3}{4}\chi''\!+\!3A'\chi'\!-\!\frac{5}{4}V_\chi\right)\!\psi.\quad\label{kgchi}
\een

One can to show that Eqs. \eqref{field1}, \eqref{field2} and \eqref{kgchi} are not independent. Taking the derivative of Eq.~\eqref{field1} with respect to $y$ and using Eq. \eqref{field2} to cancel the terms proportional to $A''$, one recovers Eq. \eqref{kgchi}. In this sense, we can discard one of these equations (or any linear combination thereof) from the analysis without loss of generality. To simplify the analysis, we choose to consider the sum of Eqs. \eqref{field1} and \eqref{field2} which takes the form
\begin{equation}\label{fieldsum}
    3\varphi A''+\varphi''-\varphi'A'=-2\chi'^2-\frac32\psi\chi^{\prime 2}.
\end{equation}

On the other hand, the equations of motion for the auxiliary fields can be written as
\ben\label{kgphi}
U_\varphi=8A''+20A'^2,\qquad\quad U_\psi=\frac32\chi^{\prime 2}+5V.
\een
In a previous work \cite{Rosa:2021tei} it was shown that $U_\varphi$, $U_\psi$ and $U$ can be dealt with as independent quantities related by the chain rule
\begin{equation}
    U'=U_\varphi \varphi'+U_\psi\psi'\,.
\end{equation}
Therefore, using the Eqs. \eqref{kgphi}, we get
\begin{equation}\label{Urelfinal}
    U'=\left(8A''+20A'^2\right)\varphi'+\left(\frac{3}{2}\chi'^2+5V\right)\psi'\,.
\end{equation}
Upon these replacements, we are left with a system of three independent equations, namely the combination of the two independent components of the field equation in Eq. \eqref{fieldsum}, the equation of motion for $\chi$ given in Eq.~\eqref{kgchi}, and the chain rule for the potential $U$ given in Eq.~\eqref{Urelfinal}. Note that the equations of motion for $\varphi$ and $\psi$ given in Eq. \eqref{kgphi} have already been removed from the system via the replacement that lead to Eq. \eqref{Urelfinal}, since the quantities $U_\varphi$ and $U_\psi$ do not appear anywhere else in the system. These three equations must be solved for the six independent quantities $A$, $\varphi$, $\psi$, $\chi$, $U$ and $V$, thus consisting of an under-determined system of equations.

To determine the system and obtain solutions, one must impose three further constraints. For this purpose, we will use the first-order formalism and write
\begin{equation}\label{FOF}
    \chi^{\prime}=W_\chi\,,\qquad\qquad A^{\prime}=-\frac23W\,,
\end{equation}
\begin{equation}\label{PotV}
    V(\chi)=\frac12W_\chi^2-\frac43W^2\,,
\end{equation}
where $W_\chi=dW/d\chi$. Note that Eqs. \eqref{FOF} and \eqref{PotV} introduce three constrains on the system while also introducing an extra quantity $W$. Thus, the system remains under-determined and a single extra constraint, e.g. the explicit form of $W$, must be introduced for determination. With that, the auxiliary fields $\varphi$, $\psi$ and the potential $U$ can be obtained from the following set of differential equations
\begin{eqnarray}\label{eqpsi}
9\psi'+2\left(8W-3W_{\chi\chi}\right)\psi=0\,,
\end{eqnarray}
\begin{equation}\label{eqvarphi}
    -2\varphi W_\chi^2+\varphi''+\frac23\varphi'W+2W_\chi^2+\frac32\psi W_\chi^2=0\,,
\end{equation}
\begin{equation}\label{eqU}
U'=\frac49\left(5W^2-3W_{\chi}^2\right)\big(4\varphi'-3\psi'\big)\,.
\end{equation}
The solutions of the Eqs. \eqref{eqpsi} to \eqref{eqU} are not analytic in general. However, as we will see, the first-order formalism allows one to obtain analytical solutions of the source field.

\section{linear stability}\label{secStability}

We can investigate the linear stability in the usual way, considering small perturbations in the scalar fields and in the metric tensor. Let us consider $\chi\to\chi(y)+\delta\chi(r,y)$ and $g_{ab}\to g_{ab}(y)+\pi_{ab}(r,y)$, where $r$ represents the four-dimensional position vector and $\pi_{ab}$ is a symmetric tensor, so that $\pi_{a4}=0$ and $\pi_{\mu\nu}=e^{2A(y)}h_{\mu\nu}(r,y)$, where $h_{\mu\nu}$ is a function that satisfies the transverse and traceless (TT) conditions, i.e., $\partial^{\mu} h_{\mu\nu}=0$ and $h_{\mu}{}^\mu=0$. We can write the perturbed metric tensor as
\begin{equation}
g_{ab}=e^{2A}\big(\eta_{\mu\nu}+h_{\mu\nu}\big)dx^{\mu}dx^{\nu}-dy^2.
\end{equation}

In Ref. \cite{Bazeia:2015owa}, the linear stability of the $f(R,T)$ brane was investigated and it was shown that if $f(R,T)=f_1(R)+f_2(T)$ the perturbation in the metric tensor decouples from the perturbation in source field. Similarly, in the scalar-tensor representation we must consider $U(\psi,\varphi)=U_1(\psi)+U_2(\varphi)$ to decouple the equations. Using these conditions, we get the equation for the perturbation $h_{\mu\nu}$ as
\begin{equation}\label{eqstabili}
\left(\partial_y^2+4A^\prime \partial_y +\frac{\varphi^\prime}{\varphi}\partial_y\right)h_{\mu\nu}=e^{-2A}\Box^{(4)}h_{\mu\nu} \,,
\end{equation}
where $\Box^{(4)}$ is the four-dimensional d'Alembert operator. Let us introduce a new $z$-coordinate defined in terms of $y$-coordinate as $dz=e^{-A(y)}dy$ and also rewrite $h_{\mu\nu}$ as
\begin{equation}
 h_{\mu\nu}(r,z)=H_{\mu\nu}(z)\,\frac{e^{-3A(z)/2}}{\sqrt{\varphi(z)}}\,e^{-ik\cdot r},
\end{equation}
where $k$ is the wave number. Then, we can rewrite Eq. \eqref{eqstabili} as a Schrodinger-like equation in the form
\ben\label{schrodinger}
-\frac{d^2H_{\mu\nu}}{dz^2}+{\cal U}(z)H_{\mu\nu}= k^2H_{\mu\nu}\,,
\een
where the potential ${\cal U}(z)$ is given by
\ben\label{potschrodinger}
{\cal U}(z)=\alpha^2(z) -\frac{d\alpha}{dz}\,.
\een
Here we defined,
\begin{equation}\nonumber
    \alpha(z)=-\frac32\frac{dA}{dz}-\frac12\frac{d}{dz}\big(\ln\varphi\big)\,.
\end{equation}

It is possible to show that Eq. \eqref{schrodinger} can be factorized as $S^{\cal y} S\,H_{\mu\nu}= k^2H_{\mu\nu}$, where $S^{\cal y}=-d/dz+\alpha(z)$ and $k^2\geq 0$. In this representation, there is no state with negative energy, and thus the model is stable against tensor perturbations. We can also get the graviton-zero mode as
\ben\label{zeromode}
{H}_{\mu\nu}^{(0)}(z)=N_{\mu\nu} \sqrt{\varphi(z)}\,e^{3A(z)/2}\,,
\een
where $N_{\mu\nu}$ is a normalization constant. Transforming back the coordinate $z$ into the coordinate $y$, the stability potential $\mathcal U$ becomes
\be\label{potschrodinger2}
{\cal U}(y)\!=\!e^{2A}\!\left(\!\frac{15 A'^2}{4}\!+\!\frac{3A''}{2}\!+\!\frac{2 A' \varphi '}{\varphi }\!+\!\frac{\varphi''}{2 \varphi }\!-\!\frac{\varphi'^2}{4 \varphi^2}\right).
\ee
In order to rewrite the zero modes in terms of $y$, we can introduce the following quantity transformation
\be
h_{\mu\nu}(r,y)=\frac{e^{-2A(y)}}{\sqrt{\varphi(y)}}\xi_{\mu\nu}(y)e^{ik\cdot r}\,.
\ee
Under this transformation, Eq. \eqref{eqstabili} can be written as
\begin{equation}\nonumber
\left(\!\partial_y\!+\!2A'\!+\!\frac{\varphi'}{2\varphi}\!\right)\!\left(\!\partial_y\!-\!2A'\!-\!\frac{\varphi'}{2\varphi}\!\right)\xi_{\mu\nu}=\frac{e^{-4A(y)}}{\sqrt{\varphi(y)}}k^2\xi_{\mu\nu}.
\end{equation}
Consequently, the zero mode obtained with $k^2=0$ takes the form
\ben\label{zeromode2}
\xi_{\mu\nu}^{(0)}(y)=N_{\mu\nu} \,\sqrt{\varphi(y)}\,e^{2A(y)}\,.
\een
Note that for the zero mode to be a real function we must consider $\varphi(y)>0$. This condition will be used in the specific models studied in what follows to impose restrictions on the parameters of the models.

\section{Specific Models}\label{SpecificModels}

\subsection{$p$ - Model}\label{modeloA}

\begin{figure}[!htb]
    \begin{center}
        \includegraphics[scale=0.6]{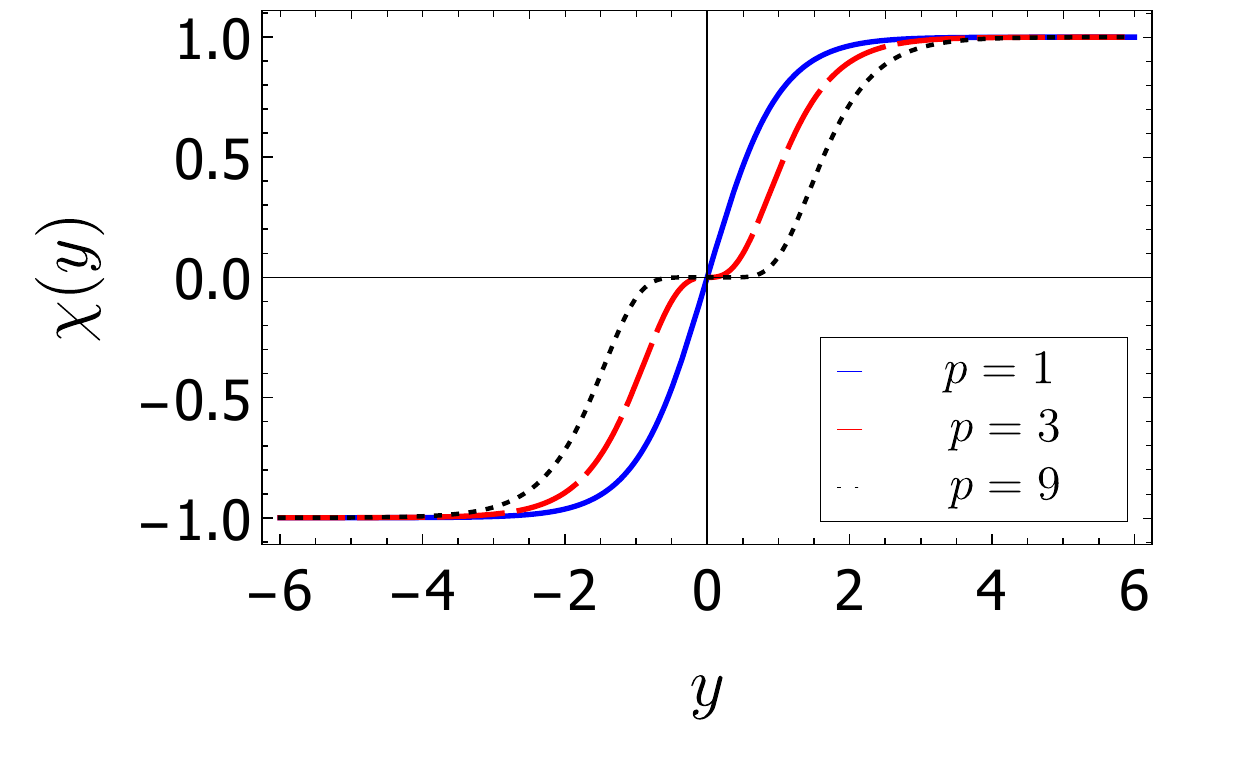}
    \end{center}
    \vspace{-0.5cm}
    \caption{\small{Profile of the source field $\chi$ given by Eq. \eqref{kinkMA} as a function of $y$.\label{fig1}}}
\end{figure}

As a first example, let us consider the so-called $p$ model initially proposed in Ref. \cite{Bazeia:2003qt} with $W(\chi)$ in the form
\begin{equation}
W(\chi)=p^2\left(\frac{\chi^{2-1/p}}{2p-1}-\frac{\chi^{2+1/p}}{2p+1}\right)\,,\label{WMA}
\end{equation}
where $p$ is a positive odd integer. As shown in what follows, this model produces a kink solution when $p=1$ and two-kink solutions when $p=3,5,7,\cdots$. The solutions for the source field $\chi$ can then be obtained from the first of the Eqs. \eqref{FOF} from which one obtains
\begin{equation}\label{kinkMA}
\chi(y)=\tanh^p(y)\,.
\end{equation}
The solutions for $\chi$ for different values of $p$ are plotted in Fig. \ref{fig1}. One verifies that for $p=1$ we have a kink-like solution which is represented by the solid line. Nonetheless, for $p=3$ and $9$, the solution becomes a two-kink solution. Furthermore, the width of the plateau region around $y=0$ increases with $p$.

\begin{figure}[!htb]
    \begin{center}
        \includegraphics[scale=0.6]{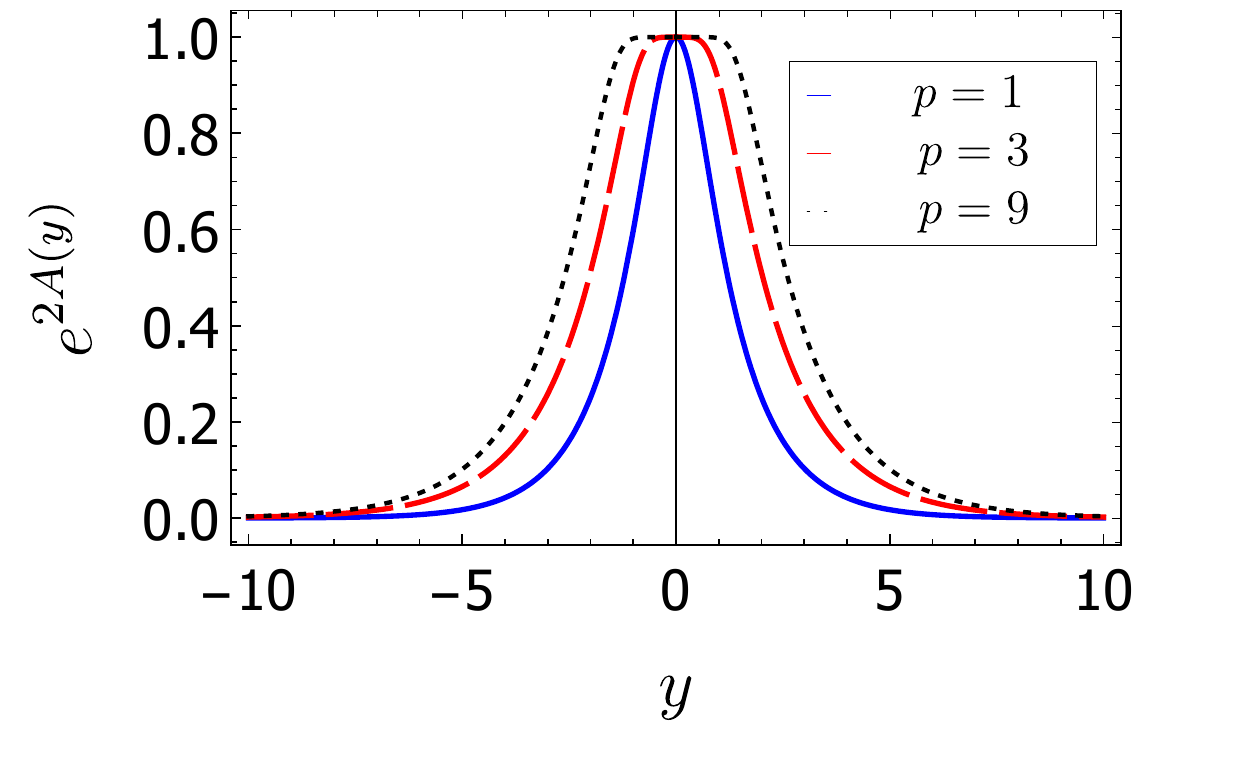}
        \includegraphics[scale=0.6]{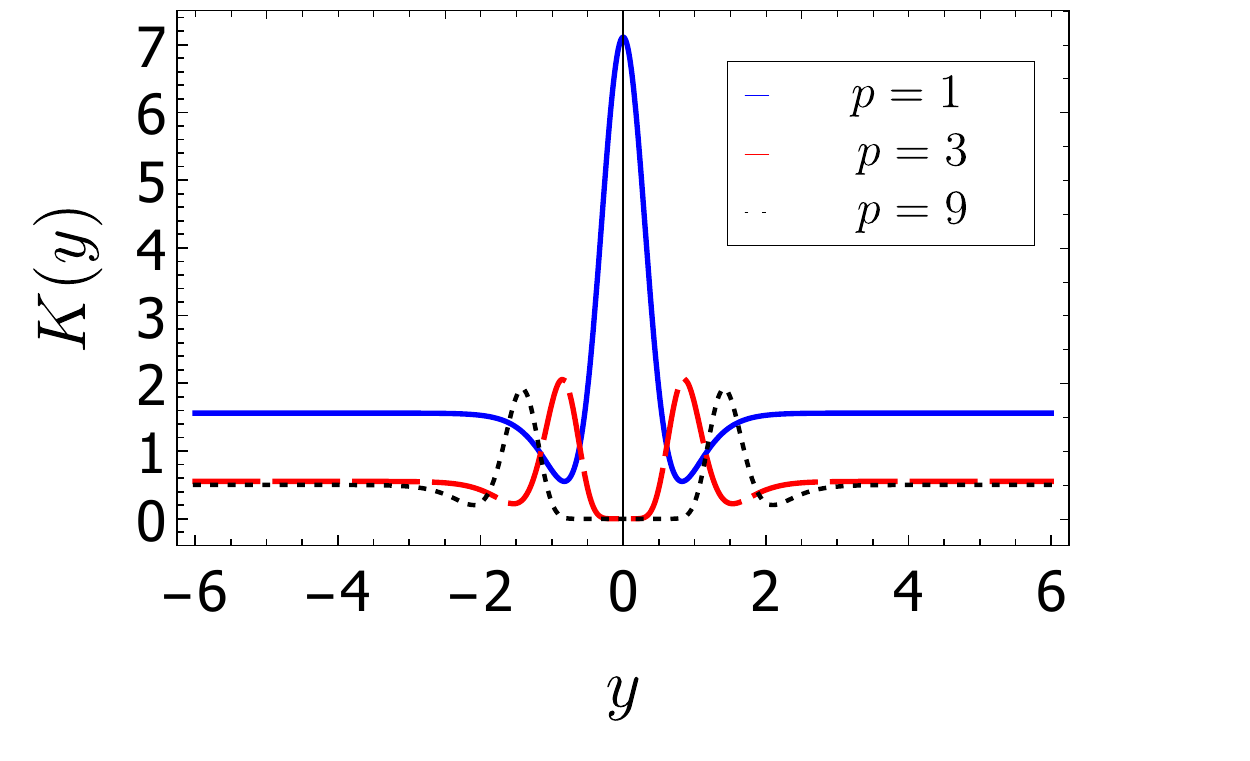}
    \end{center}
    \vspace{-0.5cm}
    \caption{\small{Warp factor (upper panel) and Kretschmann scalar (lower panel) as a function of $y$ with $A(y)$ given by Eq. \eqref{warmodA}.}\label{fig2}}
\end{figure}

The warp function $A(y)$ is analytical and can be written in terms of the hypergeometric function ${}_2F_1$ as
\ben\label{warmodA}
\!\!A(y)&=&\frac{2p\tanh^{2p}(y) }{3-12 p^2}\,{}_2F_1\!\left(1,\,p,\,1+p,\,\tanh^2(y)\right)\nn
\!\!&&-\frac{p}3\left(\frac{1-2p}{1-4p^2}\right)\tanh^{2p}(y).
\een
The upper panel of Fig. \ref{fig2} shows the warp factor $e^{2A}$ for $p=1$, $3$ and $9$. We can see that as $p$ grows the warp factor becomes wider around $y=0$.

It is interesting to verify if the spreading of the warp factor can modify the behavior of the Kretschmann scalar $K=R_{abcd}R^{abcd}$, where $R_{abcd}$ is the Riemann tensor, given by
\ben\label{KscalarMA}
K=40A'^4+16A''^2+32A'^2A''\,.
\een
Although this result is analytic, we do not write the full expression due to its size. Instead, we analyze the asymptotic behaviors of the Kretschmann scalar in $y=0$ and $y\to \pm \infty$, which are
\begin{eqnarray}\nonumber
    K(0) = \left\{
    \begin{array}{clc}
        64/9  &  \mbox{ for } &p=1\,,\\
        0 & \mbox{ for }  &p\geq 3\,,
    \end{array} \right.
\end{eqnarray}
\be\nonumber
K(y\to\pm \infty)=\frac{10240 p^8}{\left(3-12 p^2\right)^4}\,.
\ee
In the lower panel of the Fig. \ref{fig2} we plot the Kretschmann scalar for the same values of $p$ used before. We have the desired asymptotic behavior and no divergences arise. The most significant change arises in $y=0$, where for $p=1$ the scalar $K$ attains a maximum value, whereas for $p\geq 3$ it attains a minimum instead.

We can now investigate the auxiliary fields. First, we use Eqs. \eqref{eqpsi}, \eqref{WMA} and \eqref{kinkMA} to obtain the field $\psi(y)$ as
\ben\label{psimodA}
\!\!\ln\left(\frac{\psi(y)}{\psi_0}\right)&=&\frac{16 p \tanh ^{2 p}(y)}{9-36 p^2}\,{}_2F_1\left(1,\,p,\,1+p,\,\tanh ^2(y)\right)\nn
&&-\frac{8 p \tanh ^{2 p}(y)}{9+18 p}-\frac{2}{3} (1-p) \ln \left(|\tanh (y)|\right)\nn
&&+\frac{4}{3} \ln \left(\text{sech}(y)\right),
\een
where $\psi_0=\psi(0)$ is an integration constant. 
\begin{figure}[!htb]
    \begin{center}
        \includegraphics[scale=0.6]{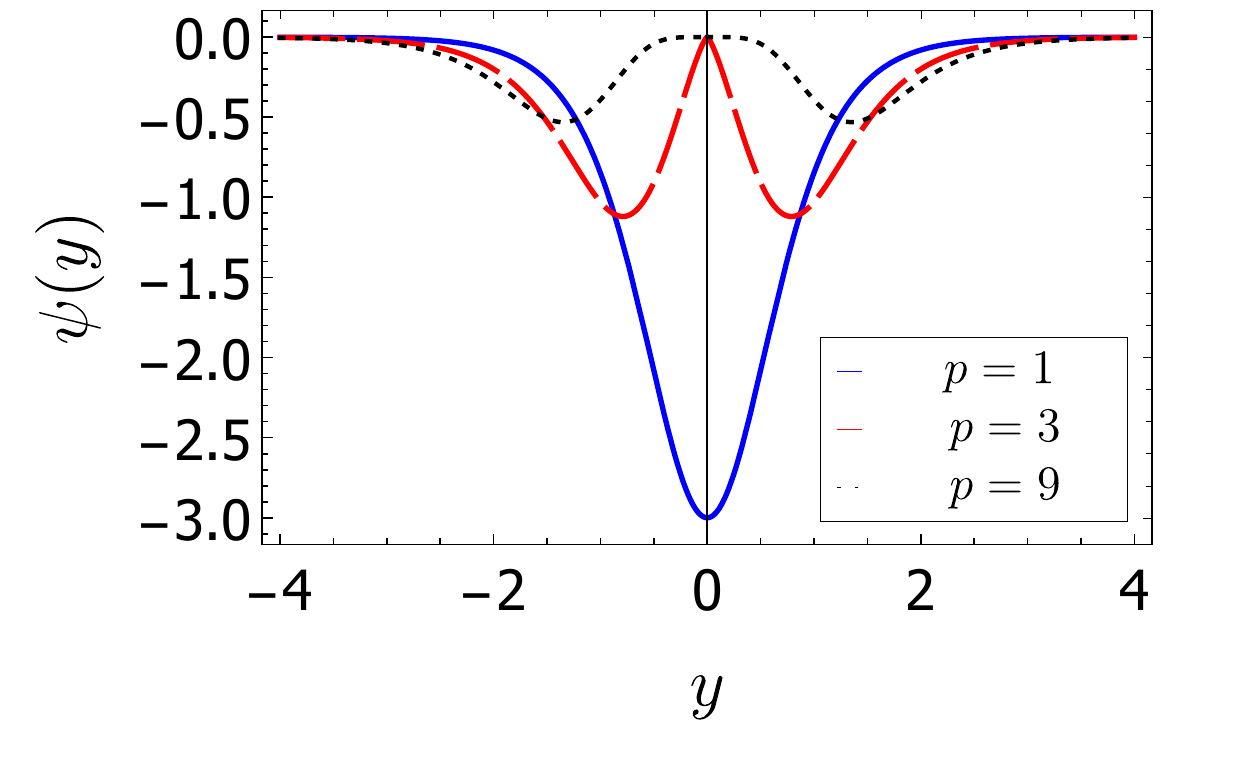}
    \end{center}
    \vspace{-0.5cm}
    \caption{\small{Auxiliary field $\psi$ given by Eq. \eqref{psimodA} as a function of $y$ for $\psi_0=-2$.}\label{fig3}}
\end{figure}
Fig. \ref{fig3} shows the auxiliary field $\psi(y)$ for $p=1$, $3$ and $9$, where we have also taken $\psi_0=-2$. Note however that the parameter $\psi_0$ functions solely as a scaling constant, see Eq. \eqref{psimodA}, and thus the value of this parameter is somewhat irrelevant for the analysis. Furthermore, note that inverting the sign of $\psi_0$ simply causes a reflection with respect to the horizontal axis. The two-kink solution appears to cause a change in the structure of the field $\psi$, which for $p\geq 3$ becomes a double peak at some $|y|=y_0\neq0$, in contrast to the single peak at $y=0$ for $p=1$. Note also that the $y_0$ at which the peaks appear increases with $p$. This result is new and different from what was previously obtained in other works, see \cite{Rosa:2021tei,Rosa:2022fhl,Rosa:2021myu}, where no internal structure has been displayed in the field $\psi$.

\begin{figure}[t]
    \begin{center}
        \includegraphics[scale=0.6]{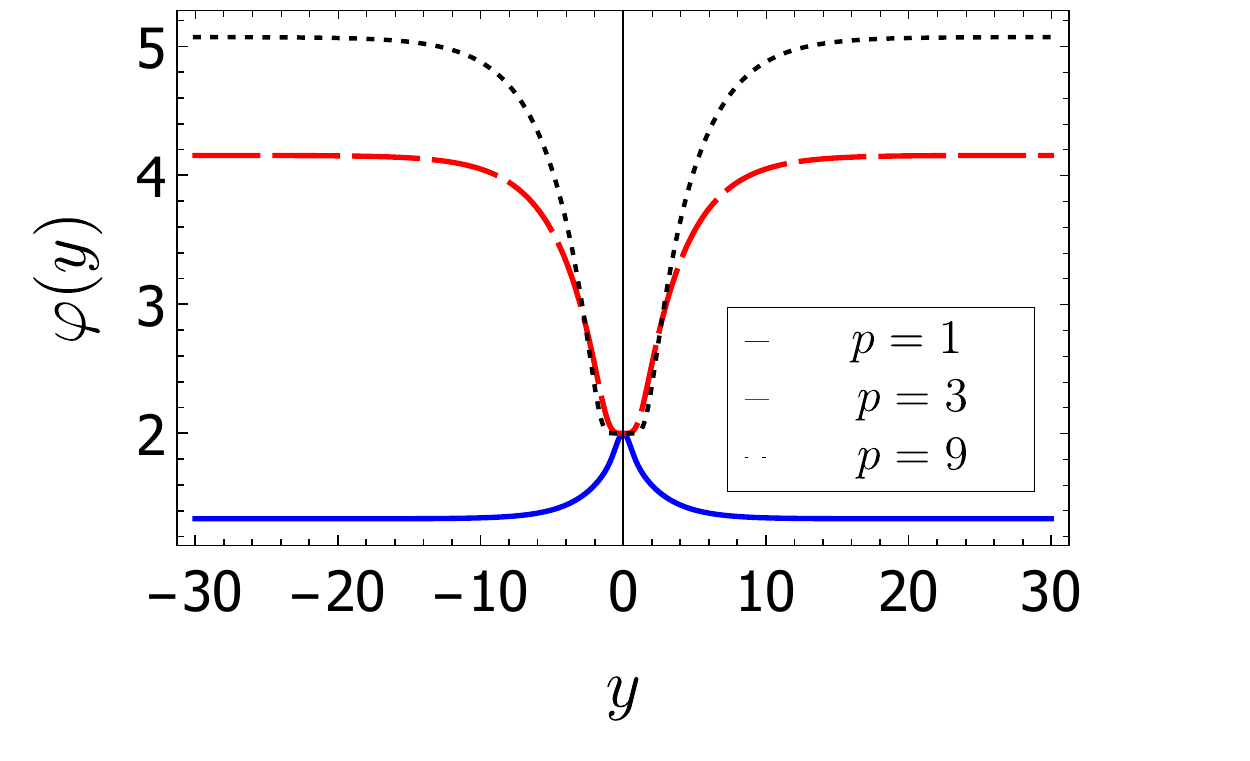}
        \includegraphics[scale=0.6]{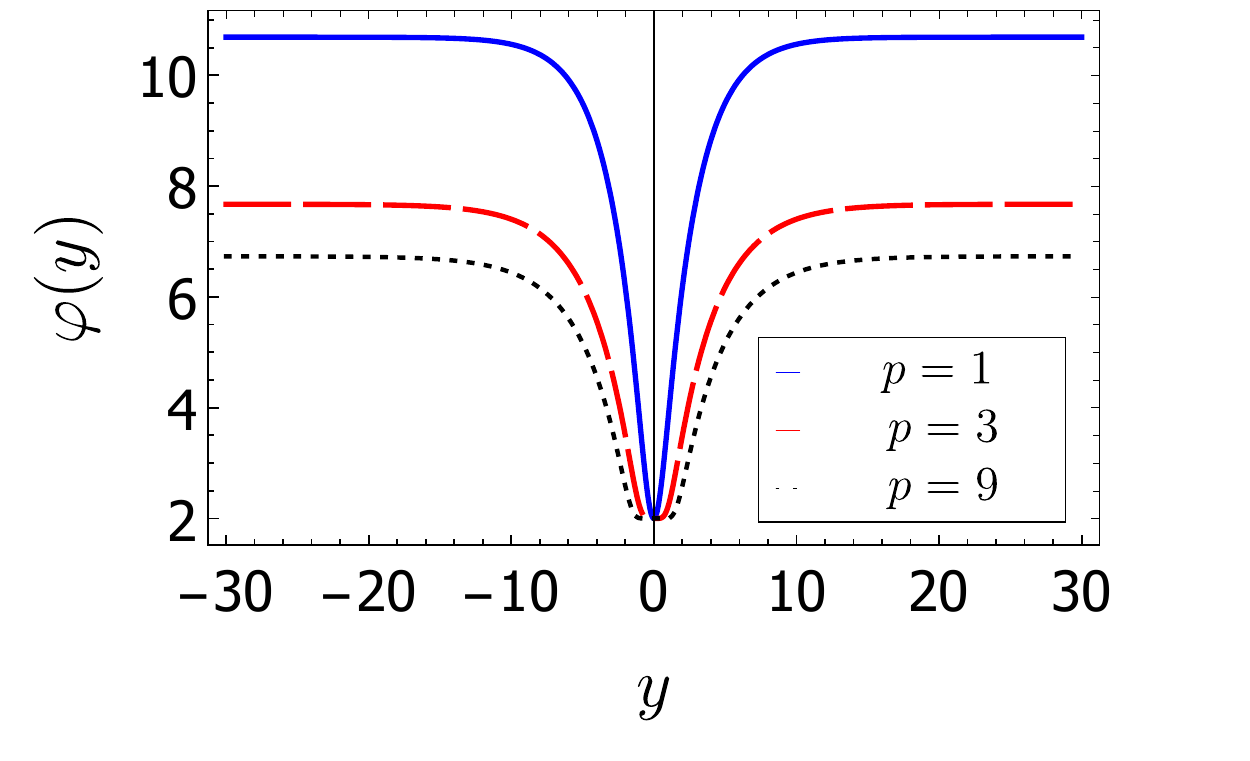}
    \end{center}
    \vspace{-0.5cm}
    \caption{\small{Numerical solutions for the auxiliary field $\varphi$ from Eq.~\eqref{eqvarphi} as a function of $y$ for $\varphi_0=2$, $\psi_0=2$ (upper panel) or $\psi_0=-2$ (lower panel).}\label{fig4}}
\end{figure}

We can now solve Eq. \eqref{eqvarphi} to obtain the solution of the auxiliary field $\varphi(y)$. However, analytical solutions for the field $\varphi$ are unattainable and we recur to numerical methods to solve this equation. Since this is a second-order differential equation, two boundary conditions must be introduced. The first boundary condition is introduced to preserve the parity of the solutions, and thus we consider $\varphi'\left(0\right)=0$. The second boundary condition is written as $\varphi\left(0\right)=\varphi_0$, for some free constant parameter $\varphi_0$. The numerical solution for $\varphi(y)$ with $\varphi_0=2$ and $\psi_0=\pm2$ is represented in Fig. \ref{fig4}. We verified that the profile of the solution is influenced by the initial conditions and by the parameter $p$ significantly, i.e., the solution may have a maximum or minimum point at $y=0$ depending on the choice of these parameters, but the general qualitative shape of the solution is not affected. Note that the only necessary restriction at this point is to guarantee that the choice of parameters preserves $\varphi>0$, which guarantees that the graviton zero mode is a real function.

\begin{figure}[t]
    \begin{center}
        \includegraphics[scale=0.6]{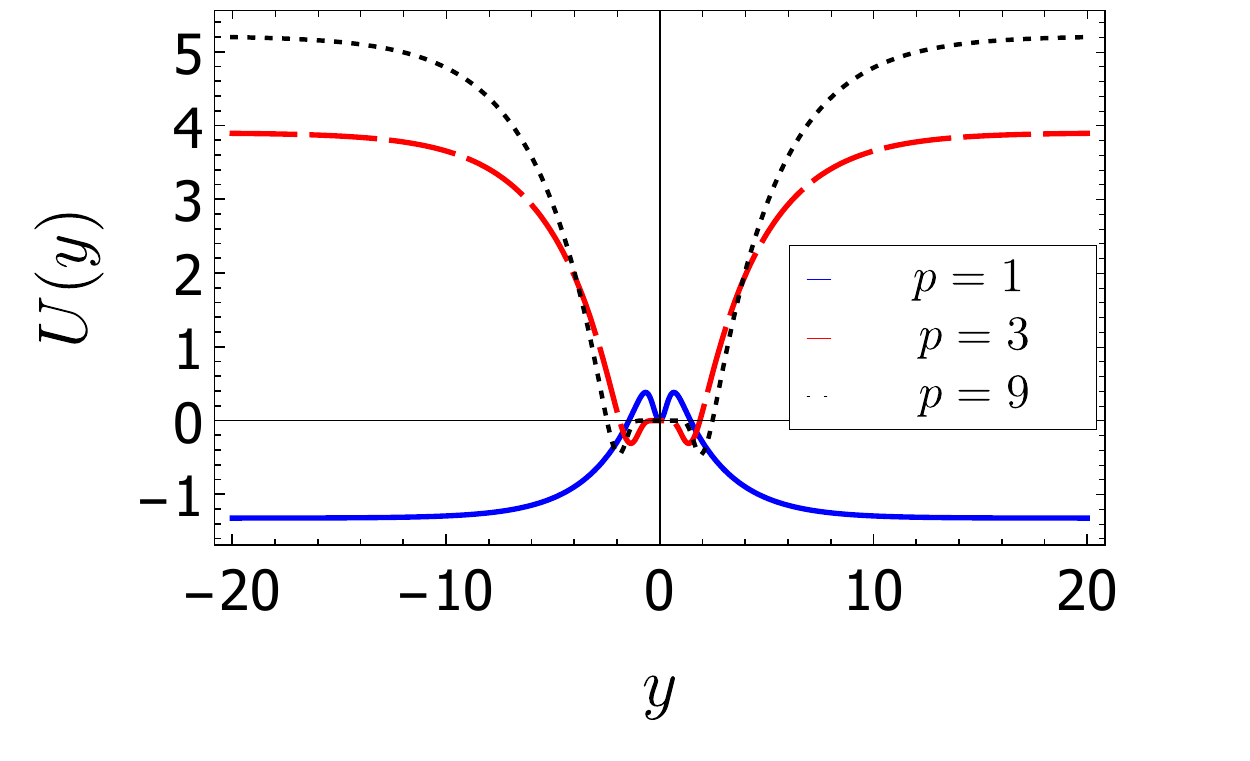}
        \includegraphics[scale=0.6]{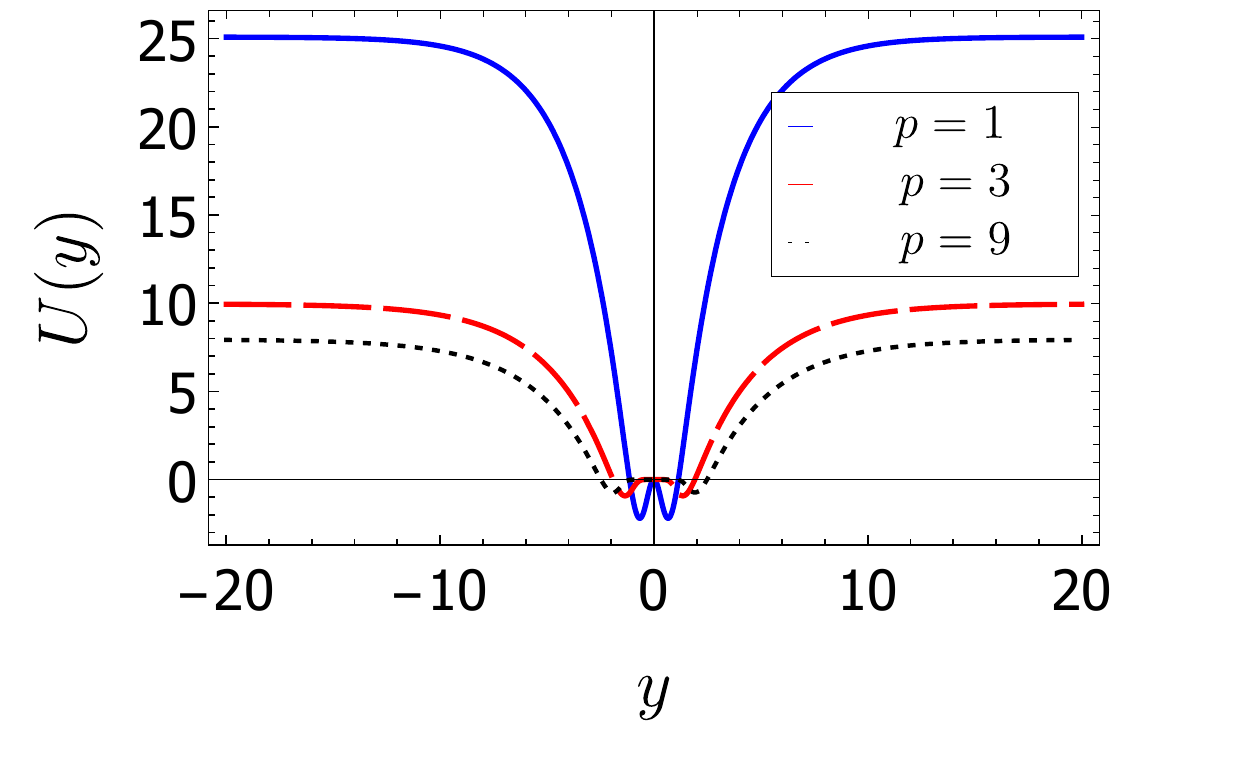}
    \end{center}
    \vspace{-0.5cm}
    \caption{\small{Numerical solutions for the potential $U(y)$ from Eq.\eqref{eqU} as a function of $y$ for $\varphi_0=2$, $\psi_0=2$ (upper panel) or $\psi_0=-2$ (lower panel).}\label{fig5}}
\end{figure}

We can also obtain the potential $U(y)$ by numerically solving Eq. \eqref{eqU} subjected to a single boundary condition $U\left(0\right)=U_0$, for some free constant parameter $U_0$. For simplicity, and since Eq. \eqref{eqU} depends solely in the derivatives of $U$, we shall take $U_0=0$ in what follows. The numerical solutions for $U\left(y\right)$ are given in Fig. \ref{fig5}. Similarly to what happens with the scalar field $\varphi$, the values of the parameters $\varphi_0$ and $p$ may affect the profile of the potential $U$, which can have either a maximum or a minimum at $y=0$, but the qualitative shape of the potential remains unaltered. Indeed, for the combinations of parameters for which $\varphi$ attains a maximum at $y=0$, the potential $U$ attains a local minimum, and vice versa.

\begin{figure}[t]
    \begin{center}
        \includegraphics[scale=0.6]{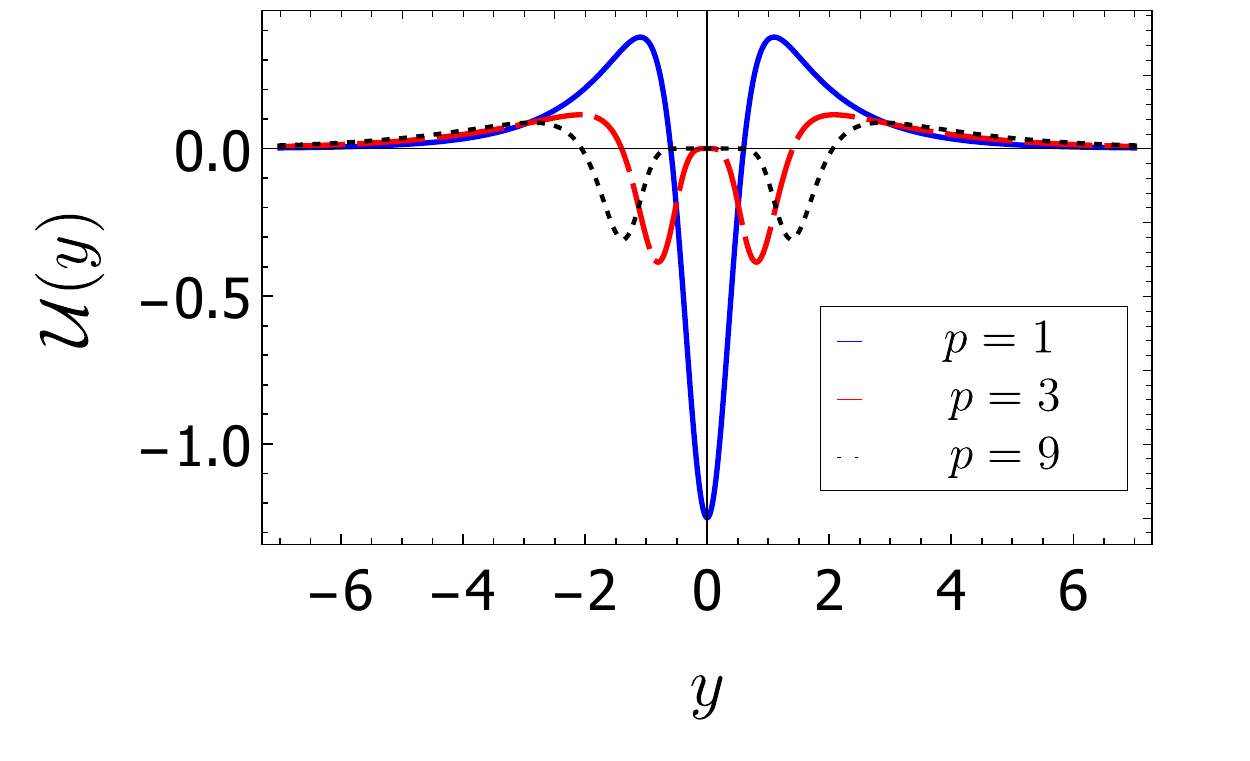}
        \includegraphics[scale=0.6]{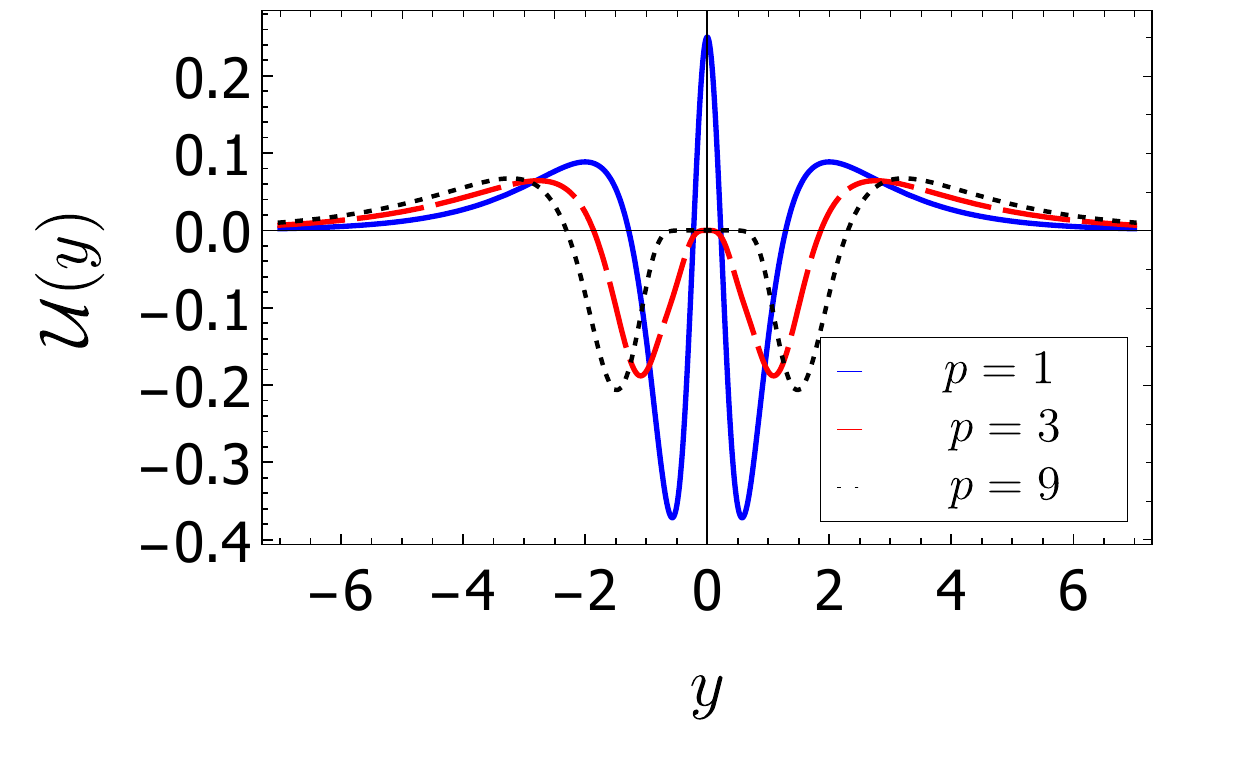}
    \end{center}
    \vspace{-0.5cm}
    \caption{\small{Stability potential ${\cal U}(y)$ from Eq. \eqref{potschrodinger2} as a function of $y$ for $\varphi_0=2$, $\psi_0=2$ (upper panel) or $\psi_0=-2$ (lower panel).}\label{fig6}}
\end{figure}

\begin{figure}[!htb]
    \begin{center}
        \includegraphics[scale=0.6]{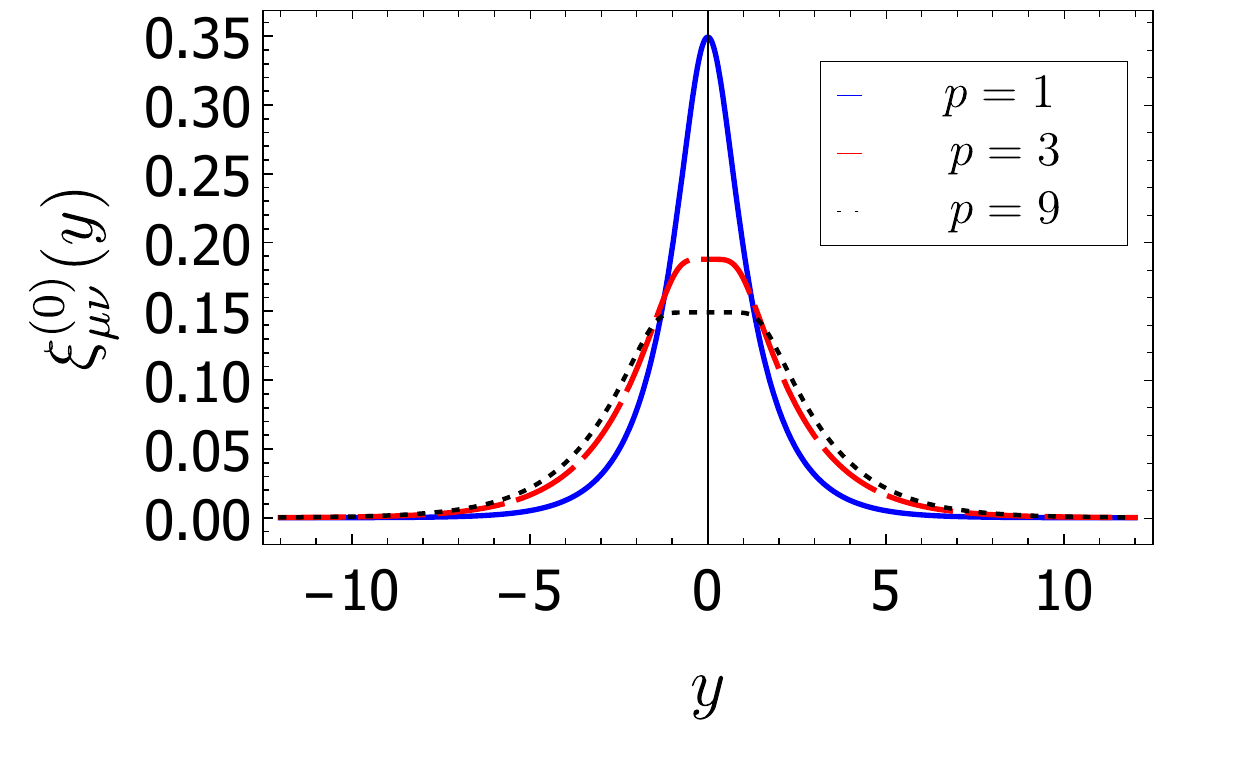}
        \includegraphics[scale=0.6]{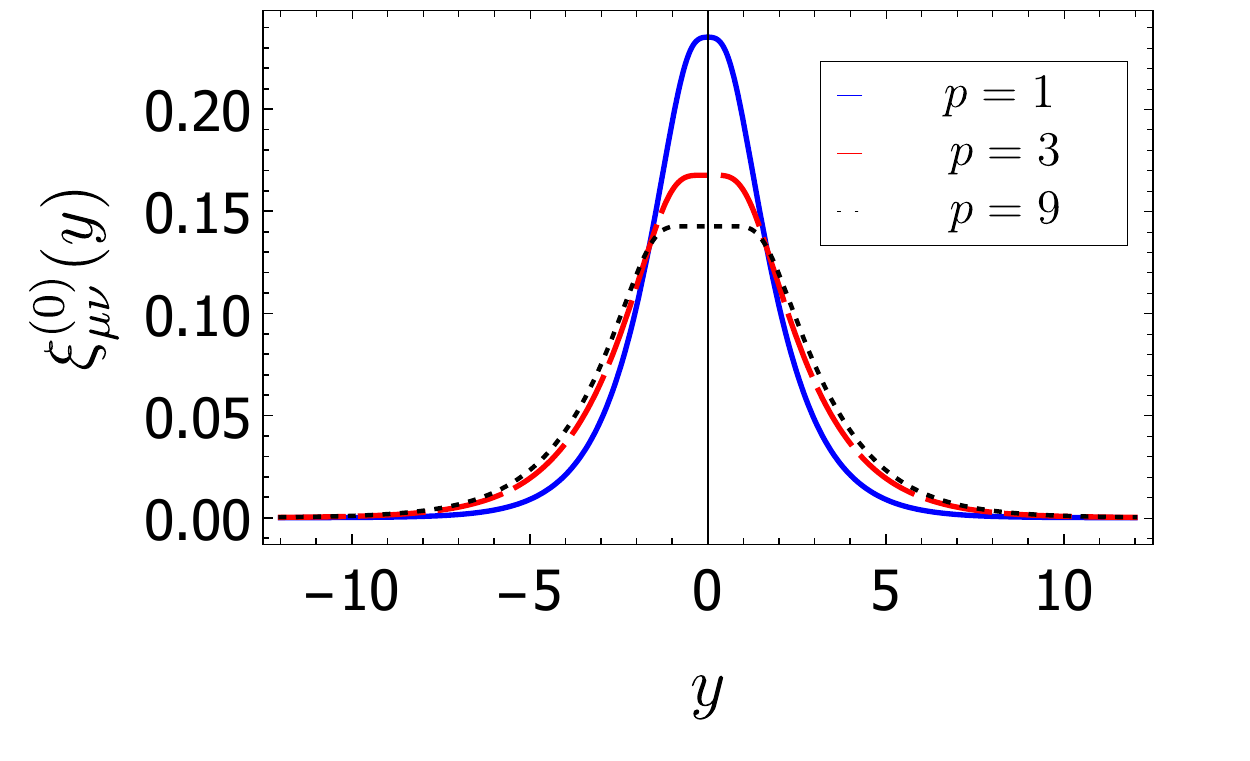}
    \end{center}
    \vspace{-0.5cm}
    \caption{\small{Zero mode $\xi^{(0)}_{\mu\nu}$ from Eq. \eqref{zeromode2} as a function of $y$ for $\varphi_0=2$, $\psi_0=2$ (upper panel) or $\psi_0=-2$ (lower panel).}\label{fig7}}
\end{figure}

The stability analysis indicates that the two-kink solution also modifies the structure of the stability potential ${\cal U}(y)$ and the zero mode $\xi_{\mu\nu}^{(0)}(y)$. These two quantities are plotted in Figs. \ref{fig6} and \ref{fig7}, respectively, for $\varphi_0=2$ and $\psi_0=2$ (upper panels) or $\psi_0=-2$ (lower panels). One verifies that for $p=1$ the potential $\cal U$ might have multiple different behaviors depending on the parameters of the model, e.g., a single potential well or a potential barrier at $y=0$. However, as $p$ is increased, the qualitative behavior of the potential $\cal U$ degenerates into a double potential-well that vanishes at $y=0$. Furthermore, the width of the plateau at $y=0$ is shown to increase with $p$, which results in a consequent flattening of the graviton zero mode. Since no potential barrier forms for $p\geq 3$, the brane does not develop an internal structure in these cases. However, an internal structure in the zero mode may arise for $p=1$, for specific choices of $\varphi_0$ and $\psi_0$, which is consistent with the fact that the stability potential presents a potential barrier at $y=0$ in this case, as also seen in \cite{Rosa:2022fhl}.  This internal structure is not visible in Figs. \ref{fig6} and \ref{fig7} due to the values of the boundary conditions $\varphi_0$ and $\psi_0$ chosen.

\subsection{Deformed model}

We now turn our attention to another interesting model that allows for kink-like solutions. Let us consider $W$ in the form
\be\label{WmodelB}
\begin{aligned}
\!\!\!\!W(\chi)=&\,\frac{a^2\left(3a^2\!-\!4\right) }{8\sqrt{(a^2\!-\!1)^3}}\,\arcsinh \left(\sqrt{\frac{1-a^2}{a^2}}\,\chi \right)\\
&+\frac{\chi}{8}\left(\frac{(4-5a^2)}{2(1-a^2)}  -\chi ^2\right)\!\sqrt{a^2\!+\!(1\!-\!a^2)\chi^2}\,.
\end{aligned}
\ee
where $a$ is a constant parameter in the range $(0,1)$. This model was constructed via a mathematical description known as the deformation method and was used in flat space to interconnect solutions of the polynomial model $\chi^4$ with solutions of the model $\chi^6$, see Ref. \cite{Bazeia:2005hu} for more details.

\begin{figure}[!htb]
    \begin{center}
        \includegraphics[scale=0.6]{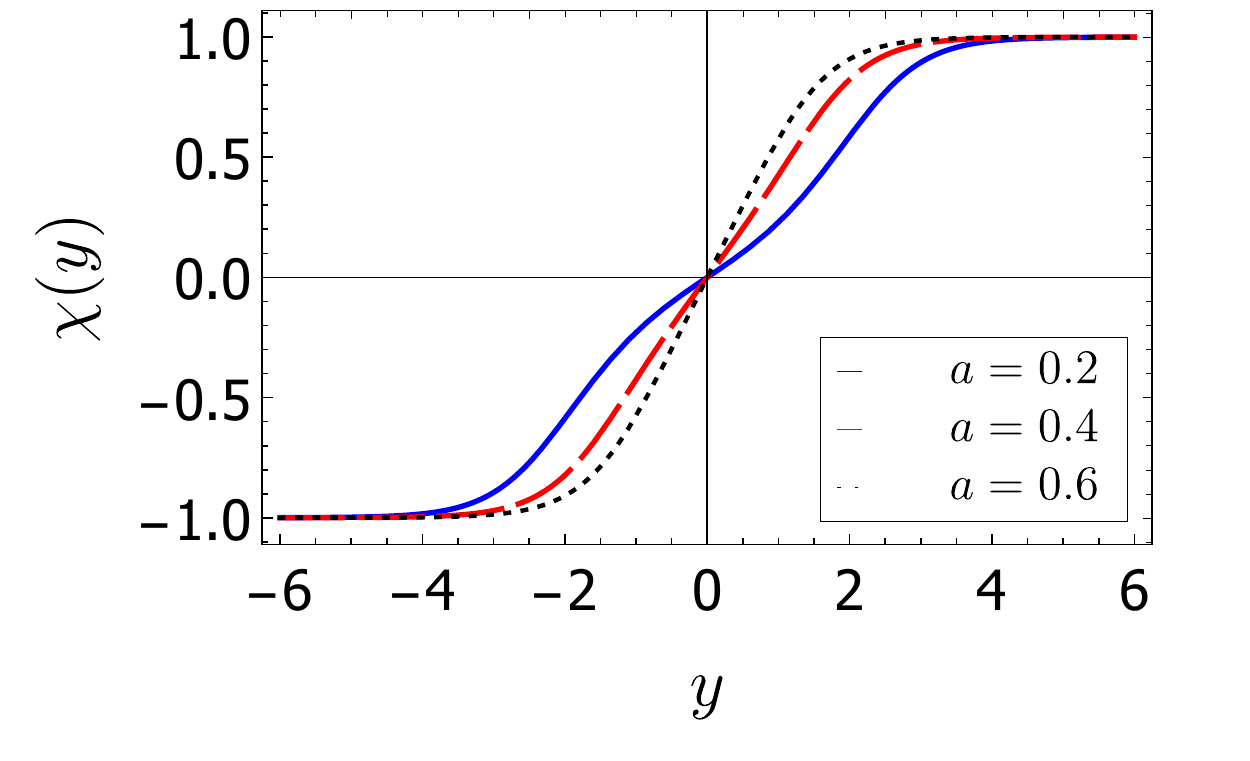}
        \includegraphics[scale=0.6]{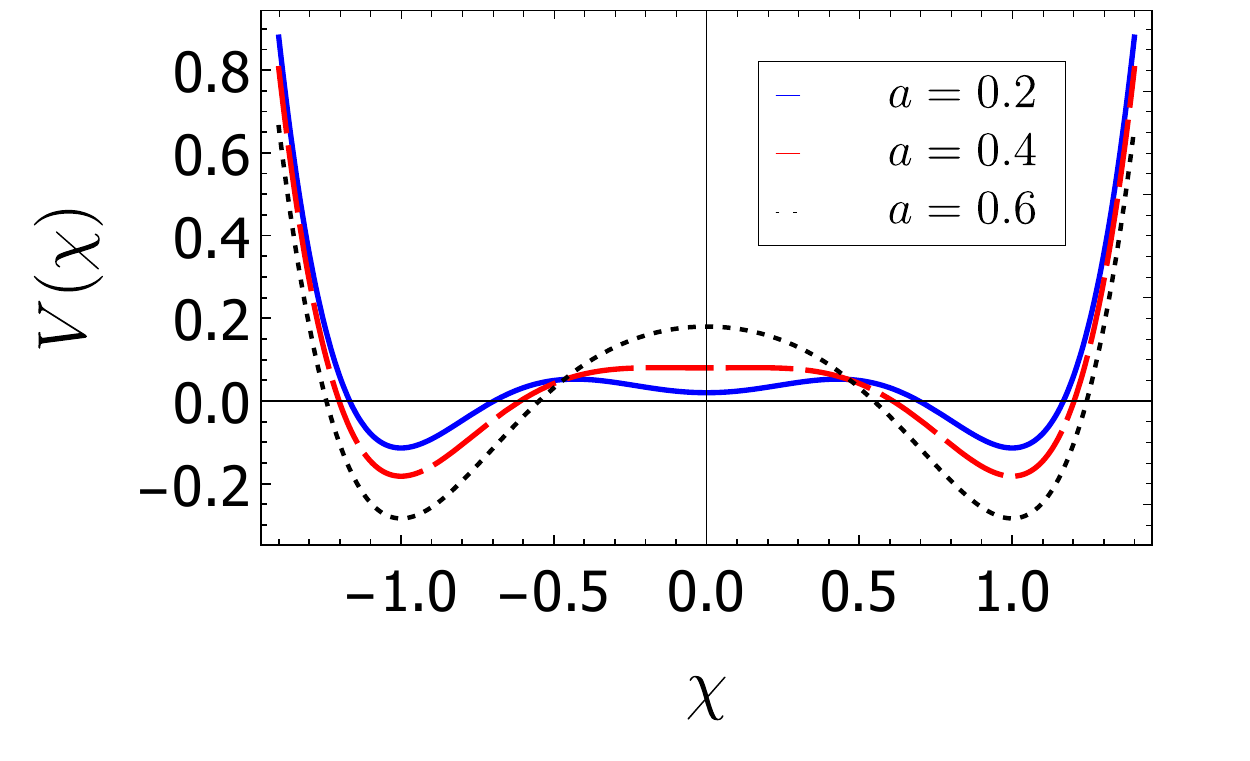}
    \end{center}
    \vspace{-0.5cm}
    \caption{\small{Profile of the source field $\chi$ given by Eq. \eqref{soluchimodb} as a function of $y$ (upper panel) and thepPotential $V$ (lower panel) as a function of $\chi$.}\label{fig8}}
\end{figure}

Using the first of Eqs. \eqref{FOF} and the $W(\chi)$ given by Eq.~\eqref{WmodelB}, we obtain the solution of the field $\chi(y)$ as
\begin{equation}\label{soluchimodb}
\chi(y)=\frac{a\left(e^{2y}-1\right)}{\sqrt{4e^{2y}+a^2\left(e^{2y}-1\right)^2}}\,.
\end{equation}
In the upper panel of Fig. \ref{fig8} we show the behaviors of the solution of the field $\chi(y)$ for $a=0.2$, $0.4$ and $0.6$. Note that the solution exhibits a transient behavior between a kink profile in the limit $a\to 1$ and a two-kink profile in the limit $a\to 0$.

Although the parameter $a$ induces a modification of the source field $\chi$ into a two-kink-like solution, there is no change in the asymptotic behaviors, which remain $\pm 1$ in the limits $y\to\pm\infty$, respectively. It is interesting to analyze the behavior of the potential $V(\chi)$ for this case. For that purpose, one substitutes $W$ from Eq. \eqref{WmodelB} into \eqref{PotV} to obtain the full expression of the potential. However, we choose to omit this equation due to its size. Instead, we plot the behavior of the potential in the lower panel of Fig. \ref{fig8}. We observe that the potential has minima points at $\chi=\pm1$ and a central point at $\chi=0$ that is a local minimum for $a<\tilde{a}$ and a local maximum $a>\tilde{a}$, where $\tilde{a}=\sqrt{3/17}\approx 0.42$. The emergence of the central minimum plays an essential role in the transition between the single-kink and the two-kink solutions.

\begin{figure}[t]
    \begin{center}
        \includegraphics[scale=0.6]{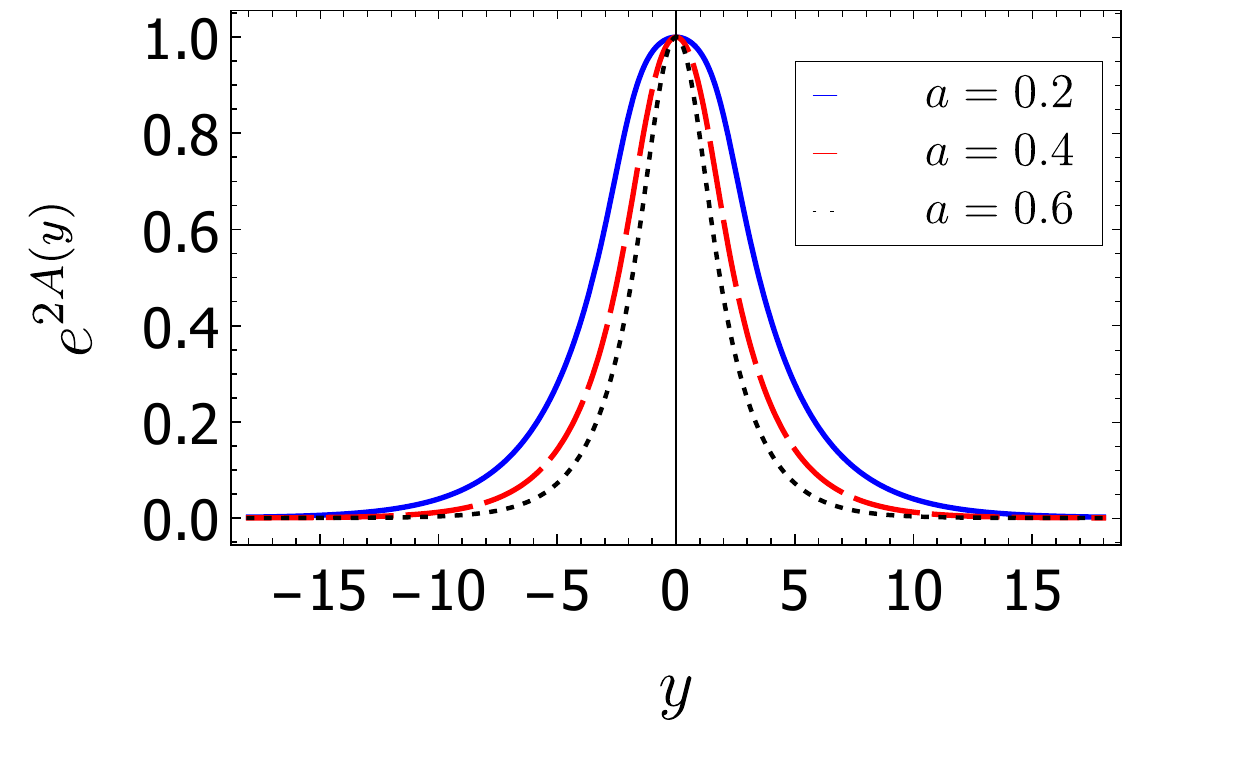}
        \includegraphics[scale=0.6]{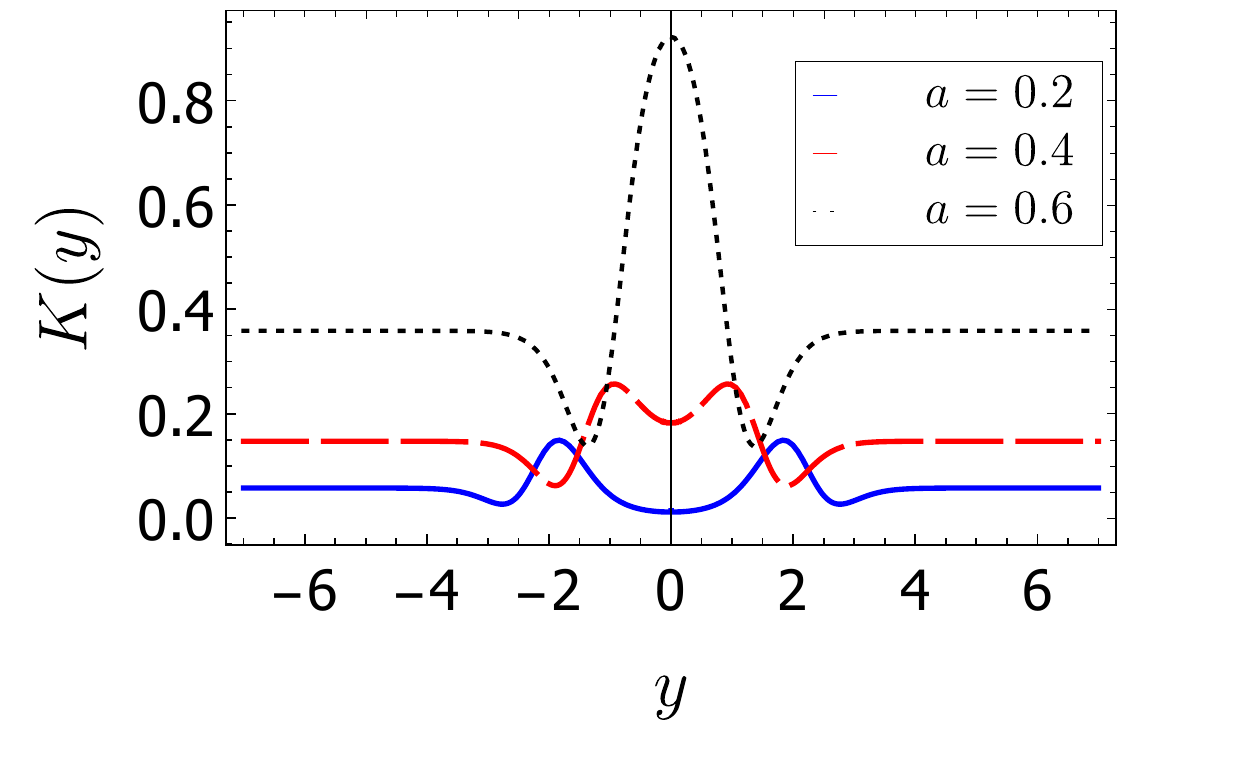}
    \end{center}
    \vspace{-0.5cm}
    \caption{\small{Warp factor (upper panel) and Kretschmann scalar (lower panel) as a function of $y$ for the deformed model in Eq.~\eqref{WmodelB}.}\label{fig09}}
\end{figure}

Let us now investigate the influence of modifying the structure of the kink solution on the behavior of the brane and the auxiliary fields of the scalar-tensor representation. We start by obtaining the warp factor via the second of Eqs. \eqref{FOF}. Given the complexity of the equations and solutions obtained so far, analytical solutions are unattainable, and we recur to numerical methods. The solutions for the warp factor are plotted in the upper panel of Fig. \ref{fig09}. We verify that when $a$ decreases to smaller and smaller values, the central region around $y=0$ is widened.

We can also verify how the Kretschmann scalar $K$ is influenced by modification of the kink solution. In the lower panel of Fig. \ref{fig09} we plot the Kretschmann scalar $K$. We verify that the regularity of $K$ is preserved, which features no divergences and asymptotically approaches constant values.

\begin{figure}[t]
    \begin{center}
        \includegraphics[scale=0.6]{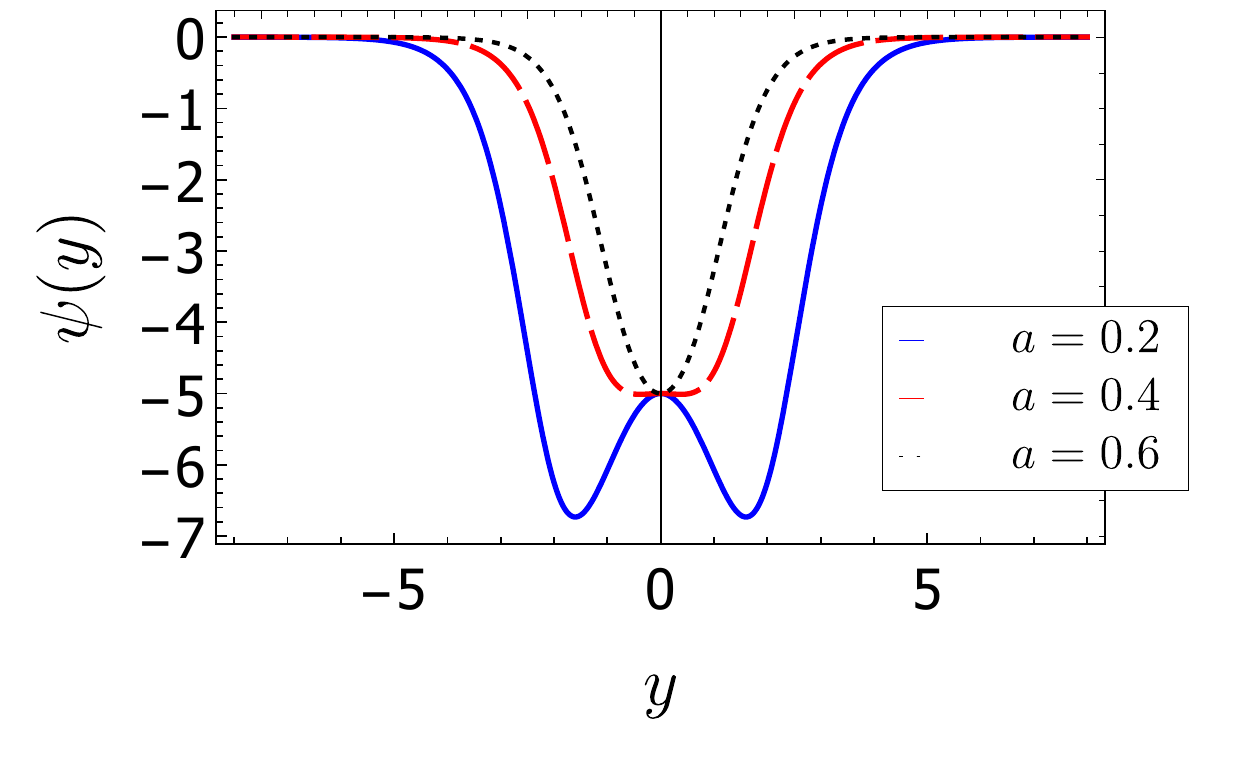}
        \includegraphics[scale=0.6]{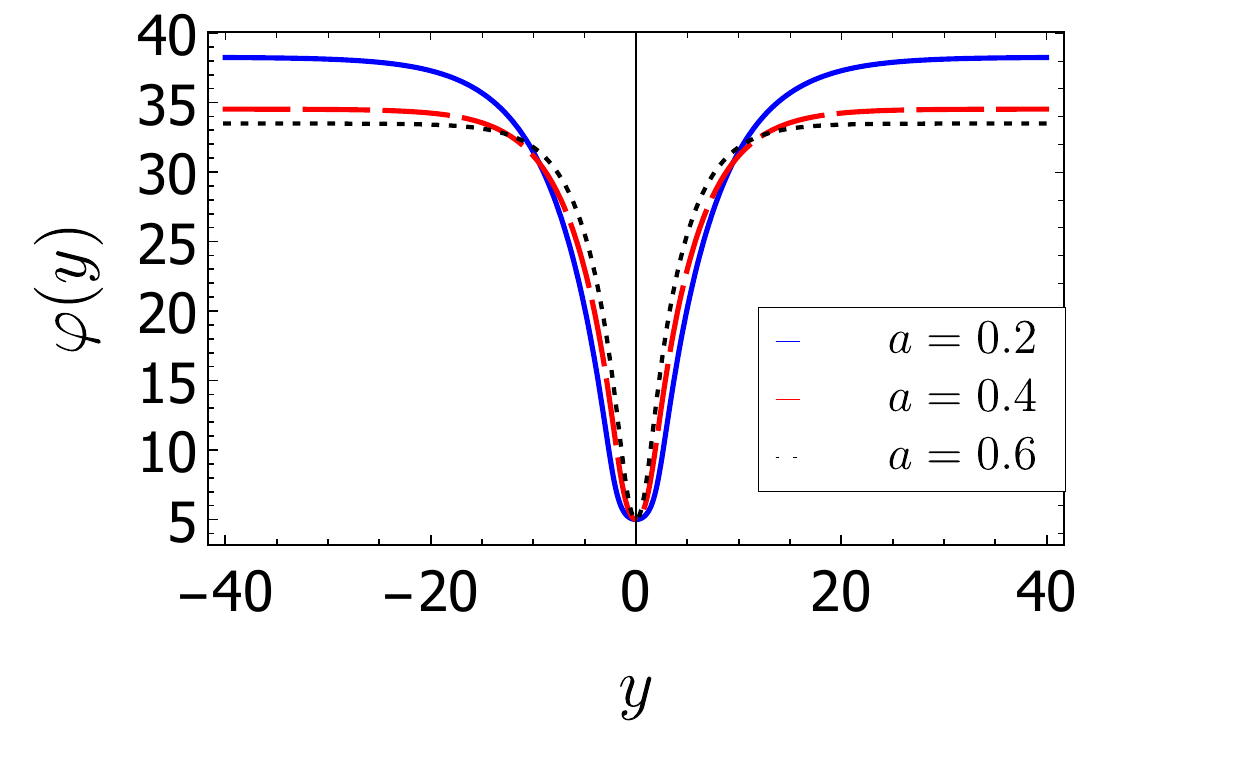}
    \end{center}
    \vspace{-0.5cm}
    \caption{\small{Numerical solutions for the auxiliary fields $\psi$ (upper panel) and $\varphi$ (lower panel) as functions of $y$ with $\varphi_0\!=\!-\psi_0\!=\!5$.}\label{fig10}}
\end{figure}

The auxiliary field $\psi(y)$ can now be obtained by numerically solving Eq. \eqref{eqpsi} subjected to a boundary condition of the form $\psi\left(0\right)=\psi_0$. The numerical solutions for $\psi$ are plotted in the upper panel of Fig. \ref{fig10} for $\psi_0=-5$. We verify that for $a<\tilde{a}$, there appears a splitting of the central minimum of the solution into two minima at $y\neq 0$ with increasing depth in the limit $a\to0$. Furthermore, we note that the value of $\psi_0$ functions as a scaling parameter, and does not affect qualitatively the behavior of the solutions. The solutions for the auxiliary field $\varphi$ can also be obtained by numerically solving Eq. \eqref{eqvarphi} subjected to the boundary conditions $\varphi\left(0\right)=\varphi_0$ and $\varphi'\left(0\right)=0$, the latter chosen to preserve the parity of the solutions. These solutions are plotted in the lower panel of Fig. \ref{fig10} where we have chosen $\varphi_0=-\psi_0=5$. Similarly to the previous model, we verify that the behavior of $\varphi$ is affected by the choice of parameters, with the solution having either a maximum or a minimum at $y=0$, but the general qualitative shape of the solution remains unaltered. Again, the only necessary restriction is to guarantee  that $\varphi>0$.

\begin{figure}[!htb]
    \begin{center}
        \includegraphics[scale=0.6]{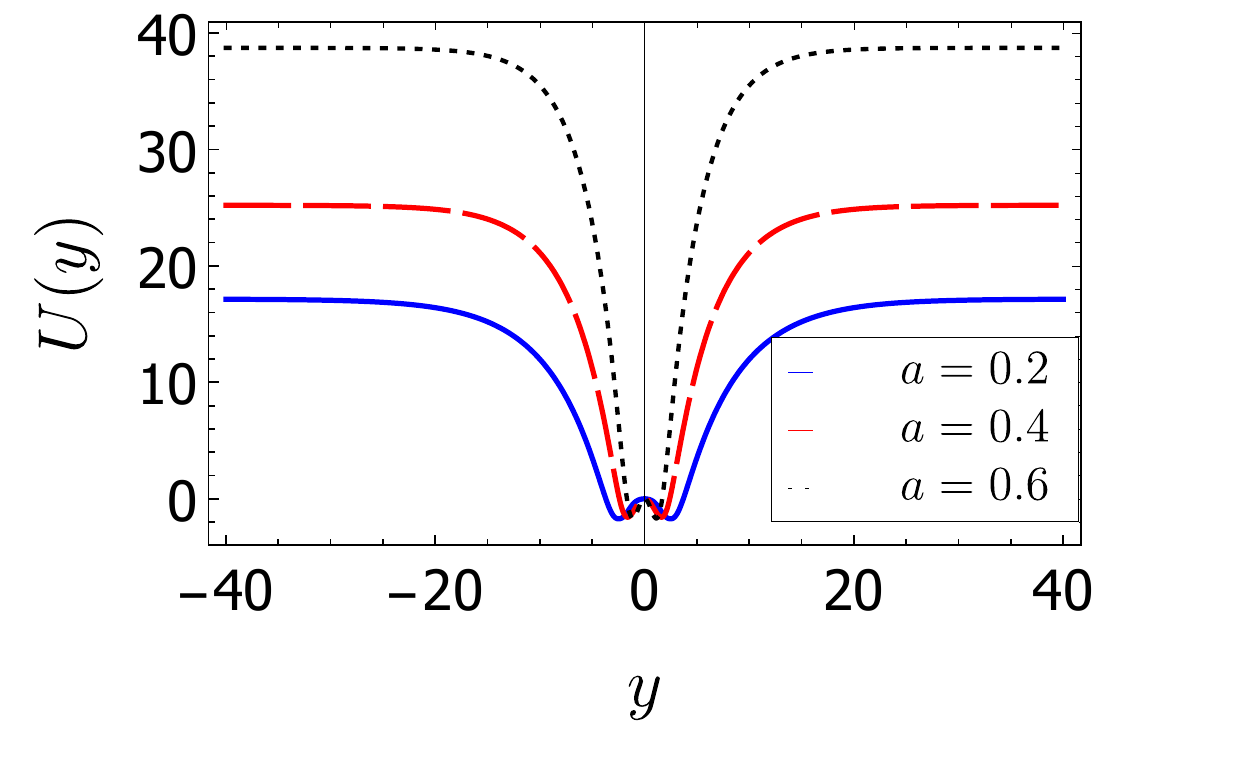}
    \end{center}
    \vspace{-0.5cm}
    \caption{\small{Numerical solutions for the potential $U$ from Eq.~\eqref{eqU} as functions of $y$ with $\varphi_0=-\psi_0=2$ and $U_0=0$.}\label{fig11}}
\end{figure}

Using the numerical solutions of the fields $\psi$ and $\varphi$ we can obtain the potential $U(y)$ by numerically solving Eq. \eqref{eqU} subjected to a boundary condition $U(0)=U_0$, where again we set $U_0=0$ without loss of generality. The solutions for $U$ are given in in Fig. \ref{fig11}. Similarly to the previous model, the shape of the potential is closely related to the shape of the scalar field $\varphi$. Indeed, the potential attains a local maximum at $y=0$ surrounded by two minima for the cases where $\varphi$ attains a minimum at $y=0$, and vice-versa.

\begin{figure}[!htb]
    \begin{center}
        \includegraphics[scale=0.6]{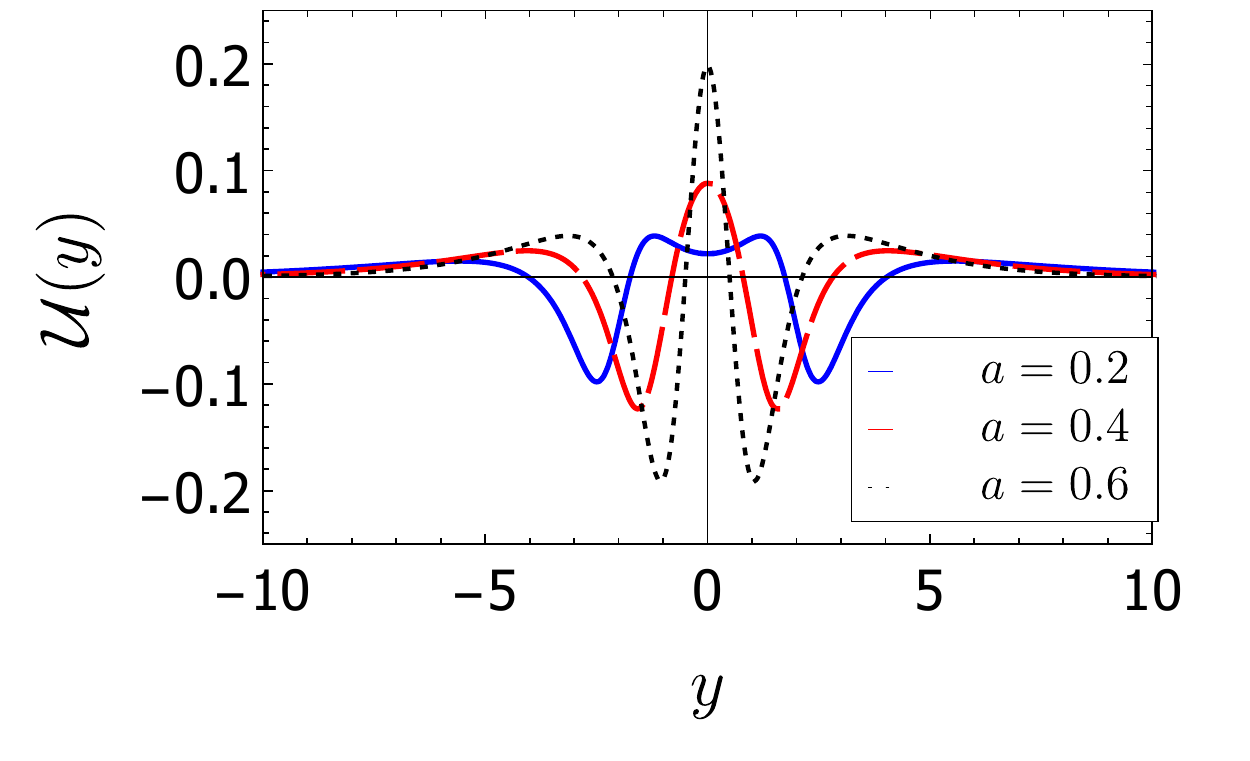}
        \includegraphics[scale=0.6]{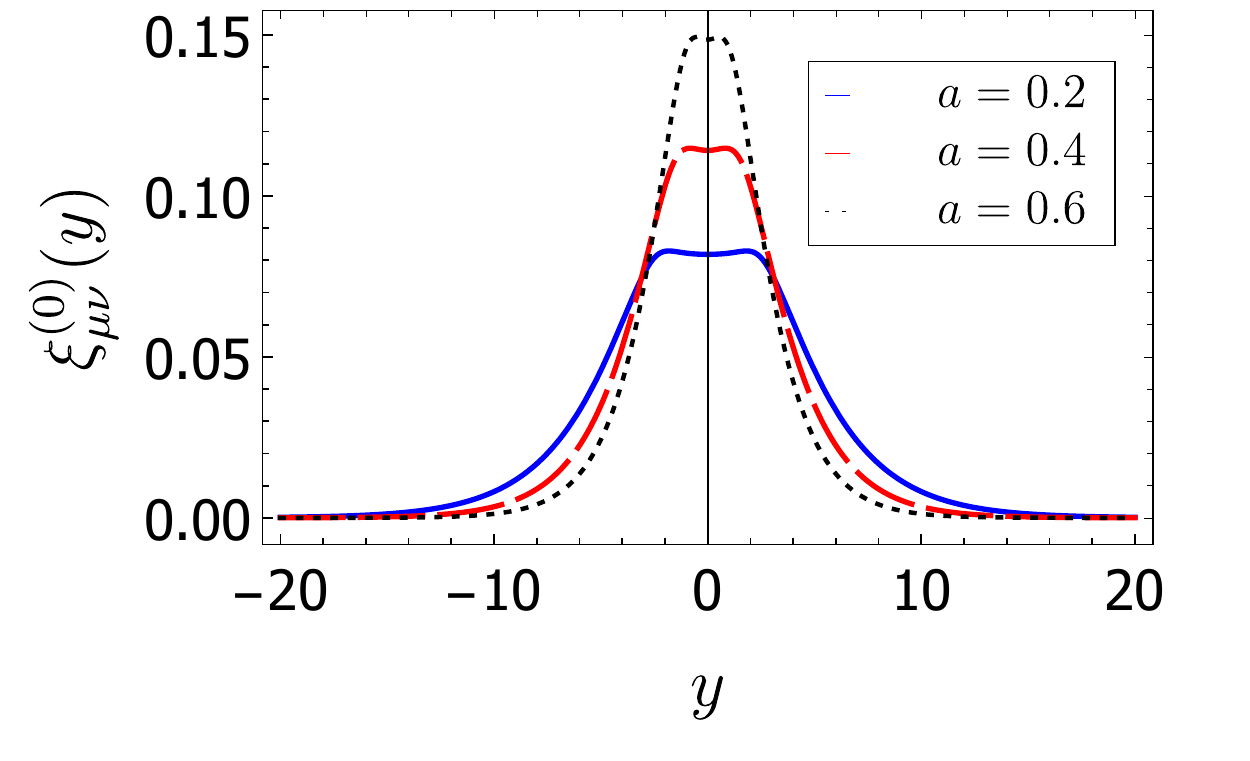}
    \end{center}
    \vspace{-0.5cm}
    \caption{\small{Stability potential (upper panel) and zero mode (lower panel) as functions of $y$ for solutions where $\varphi>0$.}\label{fig12}}
\end{figure}

Finally, in the upper panel of Fig. \ref{fig12} we plot the stability potential $\cal U$ and in the lower panel of the same figure we plot the graviton zero mode associated with the solutions for the auxiliary fields obtained in this model. It is now clear that the $a$ parameter significantly alters the behavior of the stability potential, which consequently induces an interesting modification in the internal structure of the zero mode. In addition, the stability potential presents a complex structure, which transitions between a single potential barrier at $y=0$ in the limit $a\to 1$ to a double potential barrier in the limit $a\to 0$, both situations surrounded by a double potential well.

\section{Comments and conclusions}\label{coments}

In this work, we have considered the framework of the scalar-tensor representation of $f(R,T)$ gravity to investigate models that engender multi-kink solutions for the source field of the brane. We verified how these models affect the internal structure of the auxiliary fields of the tensor-scalar representation, as well as the behavior of the stability potential and graviton zero mode.

In the first model studied, we have considered two-kink solutions obtained through the $p$-model to infer possible changes in the auxiliary fields. In this case, the solutions for the source field were adjusted by varying a discrete parameter $p$ controlling the width of the plateau at $y=0$, culminating in the emergence of an internal structure for the $\psi$ field. On the other hand, the scalar field $\varphi$ and the potential $U$ could be qualitatively affected by the parameter $p$ and by the boundary conditions at the origin chosen to numerically integrate the equations of motion. Indeed, these quantities may scale along the vertical axis but hardly develop an internal structure. The stability potential is strongly affected by the parameter $p$ and the boundary conditions, and a wide range of possible behaviors are attainable, namely a single potential barrier, a single potential well, or a double potential well. Consequently, the graviton zero mode is also strongly affected, and we have verified that an increase in $p$ leads to a flattening of the zero mode near $y=0$.

In the second model studied we investigated a situation where the single-kink is changed to a two-kink solution by the adjustment of the continuous parameter $a$. We verified that this model also induces an split in the auxiliary field $\psi$, generating the appearance of an internal structure. Similarly to the first model, we verified that the scalar field $\varphi$ does not develop internal structure and that its behavior is strongly correlated with the potential $U$. The stability potential is strongly affected by the parameter $a$, suffering a transition from a single potential barrier in the limit $a\to 1$ to a double potential barrier in the limit $a\to 0$, which consequently induces an internal structure in the graviton zero mode.

As we have seen, the emergence of multi-kink structures in the source field in braneworld models significantly influences the behavior of the auxiliary field $\psi$ of the scalar-tensor representation, changing the shape of the solutions and inducing new features. However, the auxiliary field $\varphi$ is not qualitatively affected. This effect is somewhat expected since the scalar field $\psi$ in the scalar-tensor representation carries the extra degree of freedom associated with the arbitrary dependence of the action in $T$, a quantity directly affected by the source field $\chi$. On the other hand, the scalar field $\varphi$ responds for the degree of freedom associated with the arbitrary dependence on $R$, thus being not so strongly affected by modifications in the matter sector. Furthermore, the changes induced in the stability potential (and consequently the graviton zero mode) by the multi-kink solutions may induce a modification in the resonant states generated through these models. 

The results of this work suggest the study of other possibilities, such as the Born-Infeld or the Gauss-Bonnet modification, the inclusion of extra scalar, spinorial or gauge fields to control distinct degrees of freedom. The changes presented in the above results may lead to new effects of current interest in the suggested modified braneworld scenario. Some of these novel aspects are under investigation and we hope to report on them in the near future.

\begin{acknowledgments}

DB would like to thank financial support from CNPq, grant No. 303469/2019-6. DB and ASL also thank Paraiba State Research Foundation, FAPESQ-PB, grant No. 0015/2019, for partial financial support. JLR is supported by the European Regional Development Fund and the programme Mobilitas Pluss (MOBJD647).

\end{acknowledgments}


\end{document}